\documentclass[twocolumn,aps,prl,superscriptaddress,graphicx, jmp,amsmath,amssymb,reprint]{revtex4-1} 

\usepackage{graphicx}
\usepackage[tight]{subfigure}
\usepackage{dcolumn}
\usepackage{bm}
\usepackage[mathlines]{lineno}
\usepackage{upgreek} 

\newcommand{\BABARPubYear}     {08}
\newcommand{\BABARPubNumber}  {051}
\newcommand{\SLACPubNumber}   {14947}

\newcommand{\overbar}[1]{\mkern 1.5mu\overline{\mkern-1.5mu#1\mkern-1.5mu}\mkern 1.5mu}

\input babarsym.tex

\def\BaBar{\babar\xspace}

\begin{document}

\preprint{\BaBar-PUB-\BABARPubYear/\BABARPubNumber}
\preprint{SLAC-PUB-\SLACPubNumber}

\begin{flushleft}
  \BaBar-PUB-\BABARPubYear/\BABARPubNumber \\  
 SLAC-PUB-\SLACPubNumber \\
\end{flushleft}

\begin{flushright}
\end{flushright}

\title{Study of the reaction $e^+e^-\to\jpsi\pipi$ via initial-state radiation at \babar} 

%
\author{J.~P.~Lees}
\author{V.~Poireau}
\author{V.~Tisserand}
\affiliation{Laboratoire d'Annecy-le-Vieux de Physique des Particules (LAPP), Universit\'e de Savoie, CNRS/IN2P3,  F-74941 Annecy-Le-Vieux, France}
\author{J.~Garra~Tico}
\author{E.~Grauges}
\affiliation{Universitat de Barcelona, Facultat de Fisica, Departament ECM, E-08028 Barcelona, Spain }
\author{A.~Palano$^{ab}$ }
\affiliation{INFN Sezione di Bari$^{a}$; Dipartimento di Fisica, Universit\`a di Bari$^{b}$, I-70126 Bari, Italy }
\author{G.~Eigen}
\author{B.~Stugu}
\affiliation{University of Bergen, Institute of Physics, N-5007 Bergen, Norway }
\author{D.~N.~Brown}
\author{L.~T.~Kerth}
\author{Yu.~G.~Kolomensky}
\author{G.~Lynch}
\affiliation{Lawrence Berkeley National Laboratory and University of California, Berkeley, California 94720, USA }
% ...
\collaboration{The \babar\ Collaboration}
\noaffiliation


\begin{abstract}
We study the process $e^+e^-\to\jpsi\pipi$ with initial-state-radiation events produced at the PEP-II asymmetric-energy collider. The data were recorded 
with the \BaBar detector at center-of-mass energies 10.58 and 10.54 GeV, and correspond to an integrated luminosity of  454\invfb. We investigate the $\jpsi\pipi$ mass distribution in the region from 3.5 to 5.5 \gevcc. Below 3.7 $\gevcc$ the $\psi(2S)$ signal dominates, and above 4 $\gevcc$ there is a significant peak due to the Y(4260). A fit to the data in the range 3.74 -- 5.50 \gevcc yields a mass  value $4244 ~\pm 5$ (stat) $~\pm 4$ (syst)$ \mevcc$ and a width value $114~^{+16}_{-15}$ (stat)$~\pm~7$(syst)$\mev$ for this state. We  do not confirm the report from the Belle collaboration of a broad structure at 4.01 $\gevcc$. In addition, we investigate the $\pipi$ system which results from Y(4260) decay. \\

\pacs{}{\noindent \small {PACS numbers: 13.20.Gd, 13.25.Gv, 13.66.Bc, 14.40.Pq, 12.40.Yx, 12.39.Mk, 12.39.Pn, 12.39.Ki }}

\end{abstract}
\maketitle 

\linenumbers\relax 

The observation of the $X(3872)$~\cite{ct:X3872}, followed by 
the discovery of other states such as the $\chi_{c2}(2P)(3930)$~\cite{ct:Z3930}, the $Y(3940)$~\cite{ct:Y3940}, and the $X(3940)$~\cite{ct:X3940}, has reopened interest in charmonium spectroscopy. These resonances cannot be fully explained by a simple charmonium model~\cite{ct:charmonia}. The $Y(4260)$ was discovered~\cite{ct:babar-Y}  
in the initial-state-radiation (ISR) process $e^+e^- \to \gamma_{\mathrm{ISR}}Y(4260)$, $Y$(4260)$ \to J/\psi \pi^+\pi^-$. Since it is produced directly in $e^+e^-$ annihilation, it has $J^{PC}=1^{--}$. The observation of the decay mode  $J/\psi \pi^{0}\pi^{0}$~\cite{Coan:2006rv} established that it has zero isospin.
However it is not observed to decay to $D^{*}\overbar{D}^{*}$~\cite{ct:ddstar}, nor to $D_{s}^{*}\overbar{D}_{s}^{*}$~\cite{ct:dsdstar}, so that its properties do not lend themselves to a 
simple charmonium interpretation, and its nature is still unclear. Other interpretations, such as a
four-quark state~\cite{ct:tetra,ct:tetra2}, a baryonium state~\cite{ct:baryonium}, or a hybrid state~\cite{ct:hybrid},
 have been proposed. However if the Y(4260) is a four-quark state it is expected to decay to  $D_{s}^{+}\overbar{D}_{s}^{-}$~\cite{ct:tetra2}, but this has not been observed~\cite{ct:dsdstar}.
 \indent An analysis of the reaction  $e^{+}e^{-}\to J/\psi \pi^+\pi^-$~\cite{:2007sj} which confirms the Y(4260), suggests the existence of a broad state with mass $m = 4008 \pm 40~ ^{+114}_{-28}$~\mevcc and width 
$\Gamma = 226 \pm 44~ \pm 87$~\mev.
Two additional $J^{PC}=1^{--}$ states, the $Y(4360)$ and the $Y(4660)$, have been reported in ISR production, but only in the reaction $e^{+}e^{-} \to \psi(2S) \pi^+\pi^-$~\cite{Aubert:2006ge,:2007ea}.\\
\indent In this paper we present an ISR study of the reaction
$e^{+}e^{-} \to \jpsi\pipi$ in the center-of-mass (c.m.)\ energy (E$_{\mathrm{cm}}$) range 3.5 -- 5.5 \gev. In the $J/\psi \pi^+\pi^-$ mass region below $\sim$ 3.7 $\gevcc$ the signal due to the decay $\psi(2S) \to \jpsi\pipi$ dominates. A detailed comparison to $\psi(2S)$ Monte Carlo (MC) simulation yields values of the cross section and partial width to $e^{+}e^{-}$. The high-mass tail of the $\psi(2S)$ MC distribution describes the data up to $\sim$ 4 $\gevcc$ quite well, and so we perform a maximum likelihood fit over the 3.74--5.50 $\gevcc$ mass region in which the fit function consists of the incoherent superposition of a nonresonant, decreasing exponential function describing  the $J/\psi \pi^+\pi^-$ mass region above 3.74  $\gevcc$ and a Breit-Wigner (BW) function describing production and decay of the Y(4260). Non-$J/\psi$ background is treated by means of a simultaneous  fit to the mass distribution from the $J/\psi$ sideband regions.\\
\indent This analysis uses a data sample corresponding to an integrated luminosity of 454\invfb, recorded by the \babar~detector at the
SLAC PEP-II asymmetric-energy $e^+e^-$ collider operating at c.m.\ energies 10.58 and 10.54 \gev. The detector is described in detail elsewhere~\cite{ct:babar-detector}.  
Charged-particle momenta are measured with a tracking system consisting
of a five-layer, double-sided silicon vertex tracker (SVT), and a
40-layer drift chamber (DCH), both of which are coaxial with the 1.5-T 
magnetic field of a superconducting solenoid.
An internally reflecting ring-imaging Cherenkov detector, and specific ionization measurements from the SVT and DCH,
provide charged-particle identification (PID). A CsI(Tl) electromagnetic
calorimeter (EMC) is used to detect and identify photons and electrons, and muons are identified using information from the instrumented
flux-return system.
\indent We reconstruct events corresponding to the reaction $e^+e^- \to \gamma_{\mathrm{ISR}} \jpsi\pipi$,
where $\gamma_{\mathrm{ISR}}$ represents a photon that is radiated from the 
initial state $e^{\pm}$, thus lowering the c.m.\ energy of the $e^+e^-$ collision which produces the  \jpsi\pipi system. 
We do not require observation of the ISR photon, since it is detectable in the EMC for only $\sim$ 15\% of the events. This is because the ISR photon is produced predominantly in a direction close to the $e^{+}e^{-}$ collision axis, and as such is most frequently outside the fiducial region of the EMC.\\
\indent We select events containing exactly four charged-particle tracks, and reconstruct $\jpsi$ candidates via their decay to $e^+e^-$ or $\mu^+\mu^-$.
\begin{figure}[htbp]
\centering
      \includegraphics[width=7.8cm,height=6cm]{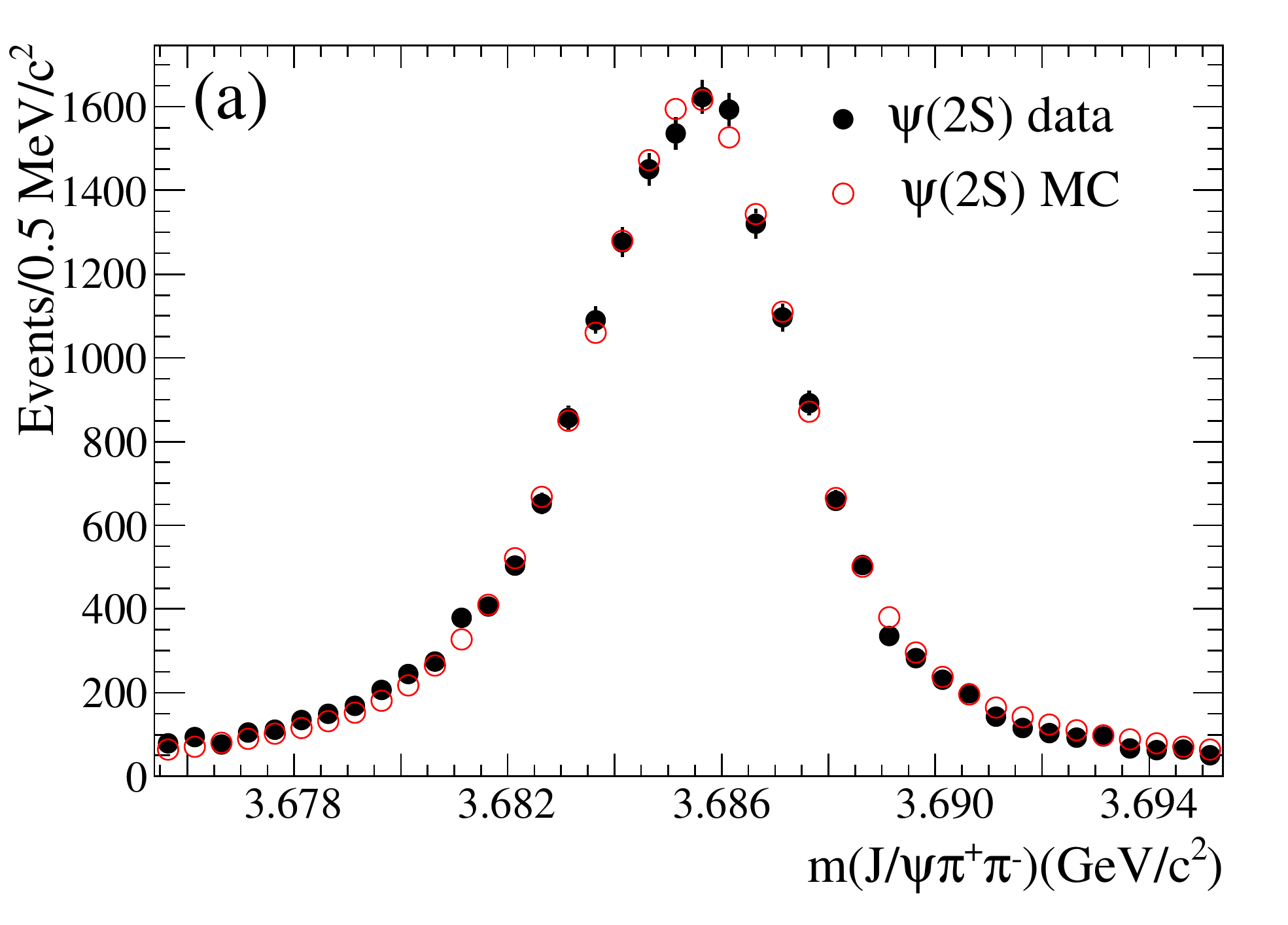} 
      \includegraphics[width=7.8cm,height=6cm]{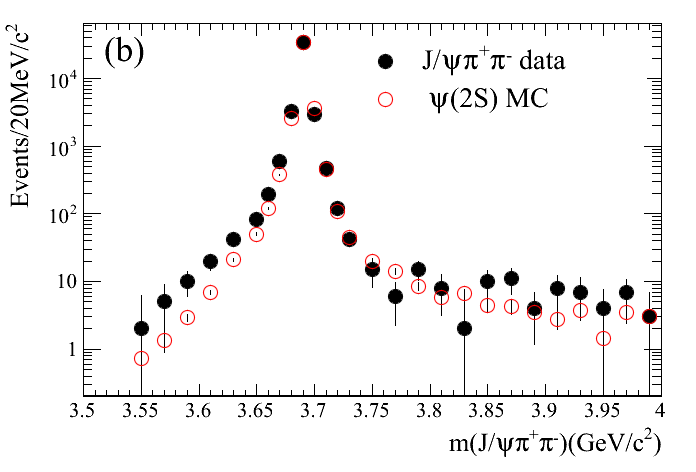}\\
     \caption{(a) The background-subtracted data, and MC simulation modified as described in the text, for the $\psi(2S)$ peak region. (b) The corresponding distribution for the mass region below 4.0 $\gevcc$.}
  \label{fig:psi2Sfig}
\end{figure}
For each mode, at least one of the leptons  must be identified on the basis of PID information.
When possible, electron candidates are combined with associated photons in order to recover bremsstrahlung energy loss, and so improve the  $\jpsi$
momentum measurement.
An $e^+e^-$ ($\mu^+\mu^-$) pair with invariant mass within ($-$75,+55)\mevcc (($-$55,+55)\mevcc) of the nominal $\jpsi$ mass~\cite{ct:PDG} is accepted as a $\jpsi$ candidate. We refer to 
the combination of these  $e^+e^-$ and $\mu^+\mu^-$ mass intervals as ``the $J/\psi$ signal region". Each $\jpsi$ candidate is subjected to a geometric fit in which the decay vertex is constrained to the $e^+e^-$ interaction region. The $\chi^{2}$ probability of this fit must be greater than $0.001$. An accepted  $\jpsi$ candidate is kinematically constrained to the nominal $\jpsi$  mass~\cite{ct:PDG} and combined with a candidate $\pi^{+}\pi^{-}$ pair in a geometric fit which must yield a vertex-$\chi^{2}$ probability greater than $0.001$.\\
\indent The value of the missing-mass-squared recoiling against the $\jpsi\pipi$ system must be in the range ($-$0.50,+0.75)~(\gevcc)$^2$ in order to be consistent with the recoil of an ISR photon. We require also that the transverse component of the missing momentum be less than 2.25\gevc. If the ISR photon is detected in the EMC, its momentum vector is added to that of the $\jpsi\pipi$ system in calculating the missing momentum. The candidate $\pi^{+}\pi^{-}$ system has a small contamination due to $e^{+}e^{-}$ pairs from photon conversions.  We compute the pair mass $m_{\mathrm{e^{+}e^{-}}}$ with the electron mass assigned to each candidate pion, and remove events with $m_{\mathrm{e^{+}e^{-}}}< 50$ \mevcc. We estimate the remaining background by using events that have an $e^{+}e^{-}(\mu^{+}\mu^{-})$ mass in the $\jpsi$ sideband ($2.896, 2.971)$ or $(3.201, 3.256)$ ($(2.936, 2.991)$ or $(3.201, 3.256)$) \gevcc after satisfying the other signal region selection criteria.\\
\indent The $\jpsi \pi^{+}\pi^{-}$ invariant-mass distribution in the region below 4 $\gevcc$ is dominated by the  $\psi (2S)$ signal. The peak region, after subtraction of background from the  \jpsi sideband regions, is shown in Fig.~\ref{fig:psi2Sfig}(a) (solid dots). The open dots indicate the $\psi(2S)$ MC distribution, modified as described below. The data distribution above $\sim$ 3.75 $\gevcc$ (Fig.~\ref{fig:psi2Sfig}(b))~may be due to the $\psi(2S)$ tail and a possible $\jpsi\pipi$ continuum ($i.e.$ nonresonant) contribution. In order to investigate this we performed a detailed comparison of the $\psi(2S)$ signal in data and in MC simulation. For the latter, we used the MC generator { \tt VECTORISR}~\cite{MC} and a simulation of the \babar~detector based on Geant4~\cite{GEANT}. The resulting MC events were subjected to the reconstruction procedures which were applied to the data. \\
\indent We first measured the peak mass position for both distributions. We performed a $\chi^{2}$-fit of a parabola to the data and MC distributions in intervals of 0.5 \mevcc 
for the region within $\pm$ 5 \mevcc of the nominal $\psi(2S)$ mass~\cite{ct:PDG}. For the data, this gave a peak mass value of 3685.32 $\pm$ 0.02~(stat) \mevcc, which is 0.77 $\pm$  0.04 \mevcc less than the nominal value~\cite{ct:PDG}. For the MC events, the result was 3685.43 $\pm$ 0.01(stat) \mevcc, which is 0.66 $\pm$ 0.01 \mevcc smaller than the input value~\cite{ct:PDG}. This difference is attributed to final-state-radiation effects. The larger deviation obtained for data may result from under-estimated energy-loss corrections, and/or magnetic field uncertainty~\cite{taumass,lambdamass}. Each MC event was then displaced by  0.11  \mevcc  toward lower mass, and the parabolic fit to the new MC distribution was repeated. The MC distribution was normalized to the data by using the data-to-MC ratio of the maxima of the fitted functions. In order to improve the MC-data resolution agreement, a $\chi^{2}$ function incorporating the data-MC histogram differences and their uncertainties was created for the region within $\pm$ 10 \mevcc of the peak mass value. In the minimization procedure each MC event was represented in mass by a superposition of two Gaussian functions with a common center, but different fractional contributions, and normalized to one event. The root-mean-squared (r.m.s.)~deviations of the Gaussian functions, and the fractional contribution of the narrower Gaussian function to the normalized distribution, were allowed to vary in the fit, and the contribution of each smeared MC event to each histogram interval was accumulated. This procedure yielded a new MC histogram to be used in the fit to the data histogram. We iterated the above procedure until the change in $\chi^{2}$ was less than 0.1, at which point the narrow (broad) Gaussian r.m.s.~deviation was 0.7 (6.3) \mevcc, and the fractional contribution was 0.88 (0.12).\\
\indent In Fig.~\ref{fig:psi2Sfig}(a) the final MC distribution is compared to the data in the fit region, and the agreement is good ($\chi^{2}/$NDF =~30.7/35, probability =~67.6 \%; NDF is the number of degrees of freedom). We integrate this MC distribution over the entire lineshape in order to estimate the $\psi(2S)$ signal yield in data, and obtain 20893 $\pm$ 145 (stat) events. We use the efficiency and the distributed luminosity (obtained from the nominal integrated luminosity and the second-order radiator function from Ref.~\cite{Kuraev:1985hb}) to obtain the cross section value 14.05 $\pm$ 0.26~(stat)~pb for radiative return to the $\psi(2S)$. This is in agreement with a previous measurement~\cite{:2007sj}. In addition we extract $\Gamma (\psi(2S) \to e^{+}e^{-}) =2.31 \pm 0.05~$(stat) keV, in excellent agreement with Ref.~\cite{ct:PDG}.\\
\indent In Fig.~\ref{fig:psi2Sfig}(b) we compare the modified $\psi(2S)$ MC distribution to the data in the region below 4.0$\gevcc$. The MC low-mass tail is systematically below the data distribution, but the high-mass tail provides a good description of the observed events. However, we note that the extrapolation to this region requires the use of the $\psi(2S)$ Breit-Wigner lineshape at mass values which are as much as 1000 full-widths beyond the central mass. The existence of many other measured final state contributions to the $J^{PC}=1^{--}$ amplitude in this mass region must call this procedure into question. Although our model adequately describes the data between the $\psi(2S)$ peak and $\sim$ 4.0 $\gevcc$, we cannot discount the possibility of a contribution from an $e^{+}e^{-} \to \jpsi \pipi$ continuum cross section in this region. In this regard, the failure of the MC lineshape to describe the data in the region of the low-mass tail might be due to the threshold rise of just such a continuum cross section. \\
\indent The $\jpsi\pipi$  mass distribution corresponding to the $\jpsi$ signal region is shown from 3.74 to 5.5 \gevcc in Fig.~2(a). The shaded histogram, which has been obtained by linear interpolation from the $J/\psi$ sideband regions, represents the estimated background contribution to the $\jpsi$ signal region.
\begin{figure}[htbp]
\centering
    \includegraphics[width=7.8cm,height=6cm]{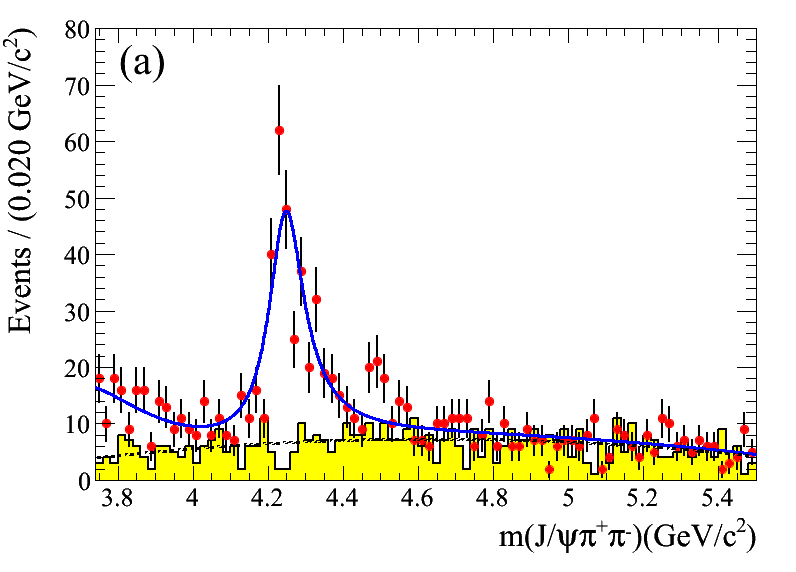}	
    \includegraphics[width=7.8cm,height=6cm]{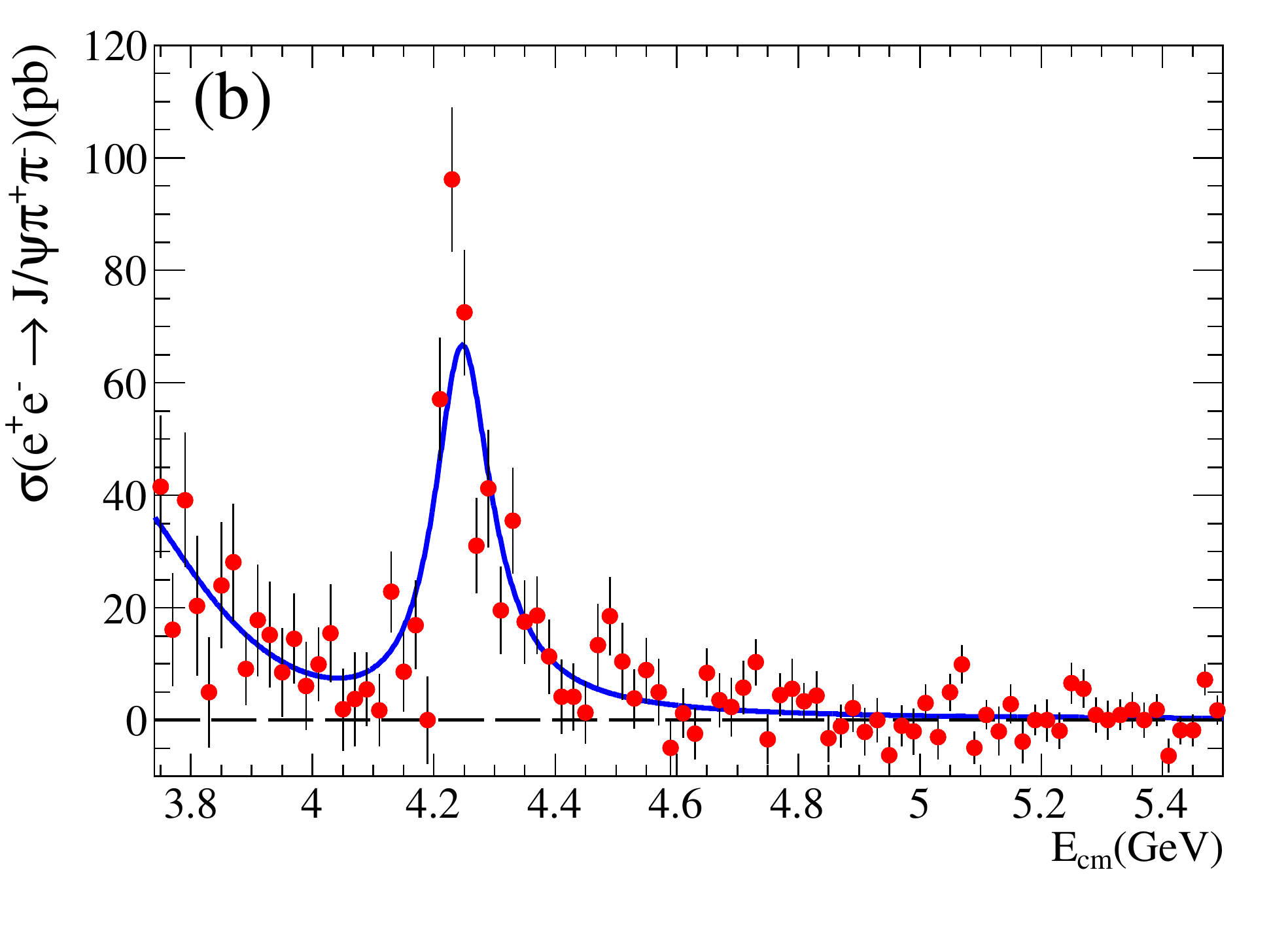}\\ 	
     \caption{(a) The $\jpsi\pipi$ mass spectrum from 3.74 \gevcc to 5.5 \gevcc; the points represent the data and the shaded histogram is the background from the  \jpsi sidebands; the solid curve represents the fit result, and the dashed curve results from the simultaneous fit to the background; (b) the measured $e^{+}e^{-} \to \jpsi\pipi$  cross section as a function of c.m.\ energy;  the solid curve results from the fit shown in (a).}
  \label{fig:fit}
\end{figure}
The signal distribution shows an excess of events over background above 3.74 $\gevcc$ which might result from the $\psi(2S)$ tail and a possible $\jpsi\pipi$ continuum contribution, as discussed with respect to Fig.~\ref{fig:psi2Sfig}(b). At higher mass we observe clear production of the Y(4260), and beyond $\sim$ 4.8 \gevcc the data are consistent with background only. There is a small excess of events near 4.5 \gevcc, which we choose to attribute to statistical fluctuation. In this regard, we note that no corresponding excess is observed in Ref.~\cite{:2007sj}. The background contribution is featureless throughout the mass region being considered.\\
\indent In order to extract the parameter values of the Y(4260), we perform an unbinned, extended-maximum-likelihood fit in the region 3.74--5.5 \gevcc to the $\jpsi\pipi$ distribution
from the $\jpsi$ signal region, and simultaneously to the background distribution from the $\jpsi$ sidebands. The background is fitted using a third-order polynomial in $\jpsi\pipi$ mass, $m$.  The mass-dependence of the signal function is given by $ f(m) = \epsilon (m) \cdot{\cal L}(m)\cdot \sigma(m)$, where $ \epsilon (m)$ is the mass-dependent signal-selection efficiency from MC simulation with a $J/\psi\pipi$ phase space distribution, and  ${\cal L}(m)$ is the mass-distributed luminosity~\cite{Kuraev:1985hb}, where we ignore the small corrections due to initial-state emission of additional soft photons; $\epsilon (m)$ increases from 9.5\% at 3.74 \gevcc to 15.5\% at 5.5 \gevcc, and ${\cal L}(m)$ from 35 pb$^{-1}$/20 MeV to 61.3 pb$^{-1}$/20 MeV  over the same range. The cross section,~$\sigma(m)$, is given by the incoherent sum $\sigma(m)=\sigma_{\mathrm{NR}}(m)+\sigma_{\mathrm{BW}}(m)$, where $\sigma_{\mathrm{NR}}(m)$ is an exponential function 
 which provides an empirical description of the $\psi(2S)$ tail and possible continuum contributions; $\sigma_{\mathrm{BW}}(m)$ is the cross section for the production of the Y(4260), and is given by \begin{equation}\label{bw}      
\begin{split}
\sigma_{\mathrm{BW}}(m) &= \frac{12 \pi C}{m^{2}} \cdot \frac{PS(m)}{PS(m_{Y})}\cdot \\
                           &  \frac{\Gamma_{e^{+}e^{-}}\cdot {\cal B}(J/\psi\pi^{+}\pi^{-}) \cdot  m_{Y}^{2}\cdot \Gamma_{Y}}{(m_{Y}^{2}-m^{2})^{2}+m_{Y}^{2}\Gamma_{Y}^{2}},
\end{split}
 \end{equation}
where $m_{Y}$ and  $\Gamma_{Y}$ are the mass and width of the Y(4260), $\Gamma_{e^{+}e^{-}}$ is the partial width for $Y(4260) \rightarrow e^{+}e^{-}$, ${\cal B}(J/\psi\pi^{+}\pi^{-})$ is the branching fraction for $Y(4260) \rightarrow J/\psi\pi^{+}\pi^{-}$, and $C = 0.3894 \times 10^{9}$ GeV$^{2}$ pb. The function $PS(m)$ represents 
the mass dependence of  $J/\psi\pi^{+}\pi^{-}$ phase space, and $PS(m_{Y})$ is its value at the mass of the Y(4260). In the likelihood function, $\sigma_{\mathrm{BW}}(m)$ is multiplied by ${\cal B}(J/\psi \to l^{+}l^{-})$, the branching fraction sum of the $e^{+}e^{-}$ and $\mu^{+}\mu^{-}$ decay modes~\cite{ct:PDG}, since the fit is to the observed events. In the fit procedure $f(m)$ is convolved with a Gaussian resolution function obtained from MC simulation. This function has a~r.m.s.~deviation which increases linearly from 2.1\mevcc at $\sim$ 3.5 \gevcc  to 5\mevcc at $\sim$ 4.3  \gevcc.
The results of the fit are shown in Fig.~\ref{fig:fit}(a). The parameter values obtained for the Y(4260) are
$m_Y = 4244 \pm 5$~(stat)$\mevcc$, $\Gamma_Y = 114^{+16}_{-15}$~(stat)$\mev$, and $ \Gamma_{e^+e^-}\times  {\cal B}(J/\psi\pi^{+}\pi^{-})  =9.2 \pm 0.8$ (stat)~eV.\\
\indent For each $J/\psi\pi^{+}\pi^{-}$ mass interval,~$i$, we calculate the $e^{+}e^{-} \rightarrow J/\psi\pi^{+}\pi^{-}$ cross section after background subtraction  using 
\begin{linenomath}
\begin{equation}\label{eq:Cross}
    \sigma_{i} = \frac{n_{i}^{\mathrm{obs}}-n_{i}^{\mathrm{bkg}}}{\epsilon_{i} \cdot {\cal L}_{i} \cdot  {\cal B}(J/ \psi \rightarrow l^{+}l^{-})} ~,
\end{equation}
\end{linenomath}
\noindent 
with $n_{i}^{\mathrm{obs}}$ and $n_{i}^{\mathrm{bkg}}$ the number of observed and background events, respectively, for this interval; $\epsilon_{i}$, and ${\cal L}_{i}$ are the values of $\epsilon(m)$ and  ${\cal L}(m)$~\cite{Kuraev:1985hb} at the center of interval $i$.\\
\indent The resulting cross section is shown in Fig.~\ref{fig:fit}(b), where the solid curve is obtained from the simultaneous likelihood fit. The corresponding estimates of systematic uncertainty are due to luminosity (1\%), tracking (5.1\%),  ${\cal B}(J/ \psi \rightarrow l^{+}l^{-})$ (0.7\%), efficiency (1\%) and PID (1\%); combined in quadrature. These yield a net systematic uncertainty of 5.4\%, as indicated in Table \ref{tab:systematic} .\\
\indent The reaction $e^{+}e^{-}\to J/\psi\pi^{+}\pi^{-}$ has been studied at the c.m. energy of the $\psi(3770)$ by the CLEO~\cite{Adam:2005mr} and BES~\cite{BES} collaborations. The former reported the value 12.1 $\pm$ 2.2 pb for the $e^{+}e^{-} \to \psi(3770) \to  J/\psi\pi^{+}\pi^{-}$ cross section, after subtraction of the contribution resulting from radiative return to the $\psi(2S)$. The dependence on E$_{{\mathrm{cm}}}$ of our fitted cross section, shown by the curve in Fig. \ref{fig:fit}(b), yields the value 31 $\pm$ 5 (stat) $\pm$ 2 (syst) pb at the $\psi(3770)$ with no subtraction of a $\psi(2S)$ contribution. This is compatible with the much more precise CLEO result obtained after subtraction. No cross section value is reported in Ref.~\cite{BES}, but the results of the BES analysis agree within their significantly larger uncertainties with those from CLEO.\\
\begin{table} 
\caption{\label{tab:systematic} Systematic uncertainty estimates for the Y(4260) parameter values.} 
\begin{ruledtabular} 
\begin{tabular}{cccc} 
Source & $\Gamma_{e^+e^-}\cdot{\cal B}(\%)$ & Mass (\mevcc) & $\Gamma$ (\mev)  \\ 
\hline 
\\
Fit procedure & $^{+1.5}_{-0.5}$ & $^{+0}_{-1}$ & $^{+2}_{-1}$ \\ 
Mass Scale  & - & $\pm 0.6$ & - \\ 
Mass resolution & - & - & $\pm 1.5$ \\ 
MC dipion model & $\pm 3.6$ & - & - \\
Decay angular \\momentum  & $\pm 3.6$ & $\pm 3.5$ & $\pm 7$ \\ 
Luminosity, etc.  & $\pm 5.4 $ & - & - \\ 
(see text)  &  & &  \\ 

\\
\end{tabular} 
\end{ruledtabular} 
\end{table} 
\indent The systematic uncertainties on the measured values of the Y(4260) parameters include contributions from the fitting procedure 
(evaluated by changing the fit range and the background parametrization), the uncertainty in the mass scale, the mass-resolution function, and the change in efficiency when the dipion distribution 
is simulated using the solid histogram in Fig.~3(c),~which is described below. In Eq.~(\ref{bw}) it is assumed that Y(4260) decay to a $J/\psi$ and a scalar dipion occurs in an $S$-wave orbital angular 
momentum state. However, a $D$-wave decay between the $J/\psi$ and the $\pipi$ system can occur also, and for this hypothesis the fitted central values of mass, width, and $\Gamma_{e^{+}e^{-}} \times {\cal B}(\jpsi\pipi) $
become 4237 \mevcc, 100 MeV, and 8.5 eV, respectively. We assign half the change in central value of each quantity as a conservative estimate of  systematic uncertainty associated with the decay angular momentum. Uncertainties associated with luminosity, tracking, ${\cal B}(J/ \psi \rightarrow l^{+}l^{-})$, efficiency and PID affect only $\Gamma_{e^+e^-}\cdot{\cal B}$, and their net contribution is 5.4\%, as we discussed previously. Our estimates of systematic uncertainty are summarized  in Table~\ref{tab:systematic}, and are combined in quadrature to obtain the values which we quote for the Y(4260).\\
\begin{figure}[!h]
\centering
      \includegraphics[width=7.8cm,height=6cm]{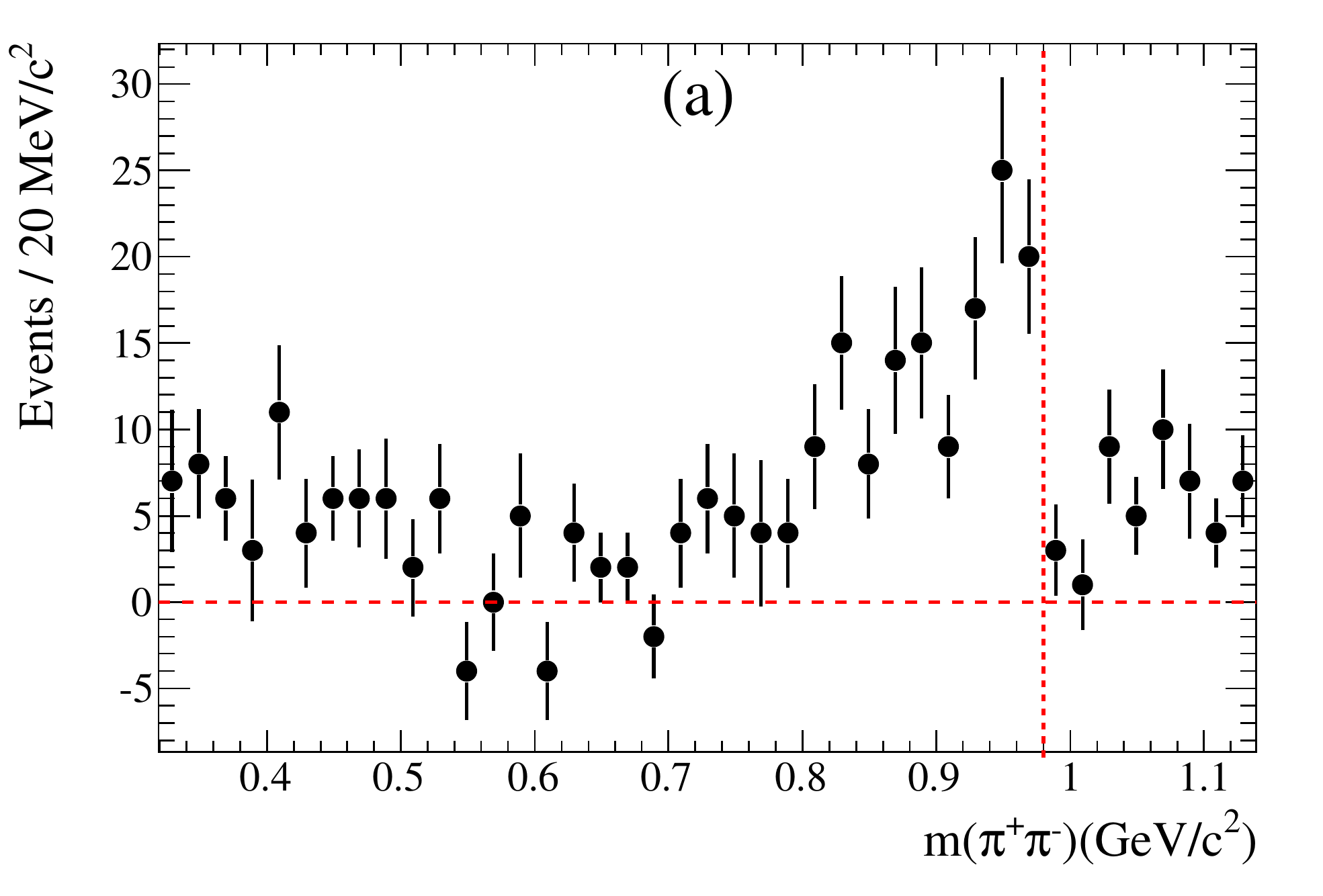}
     \includegraphics[width=7.8cm,height=6cm]{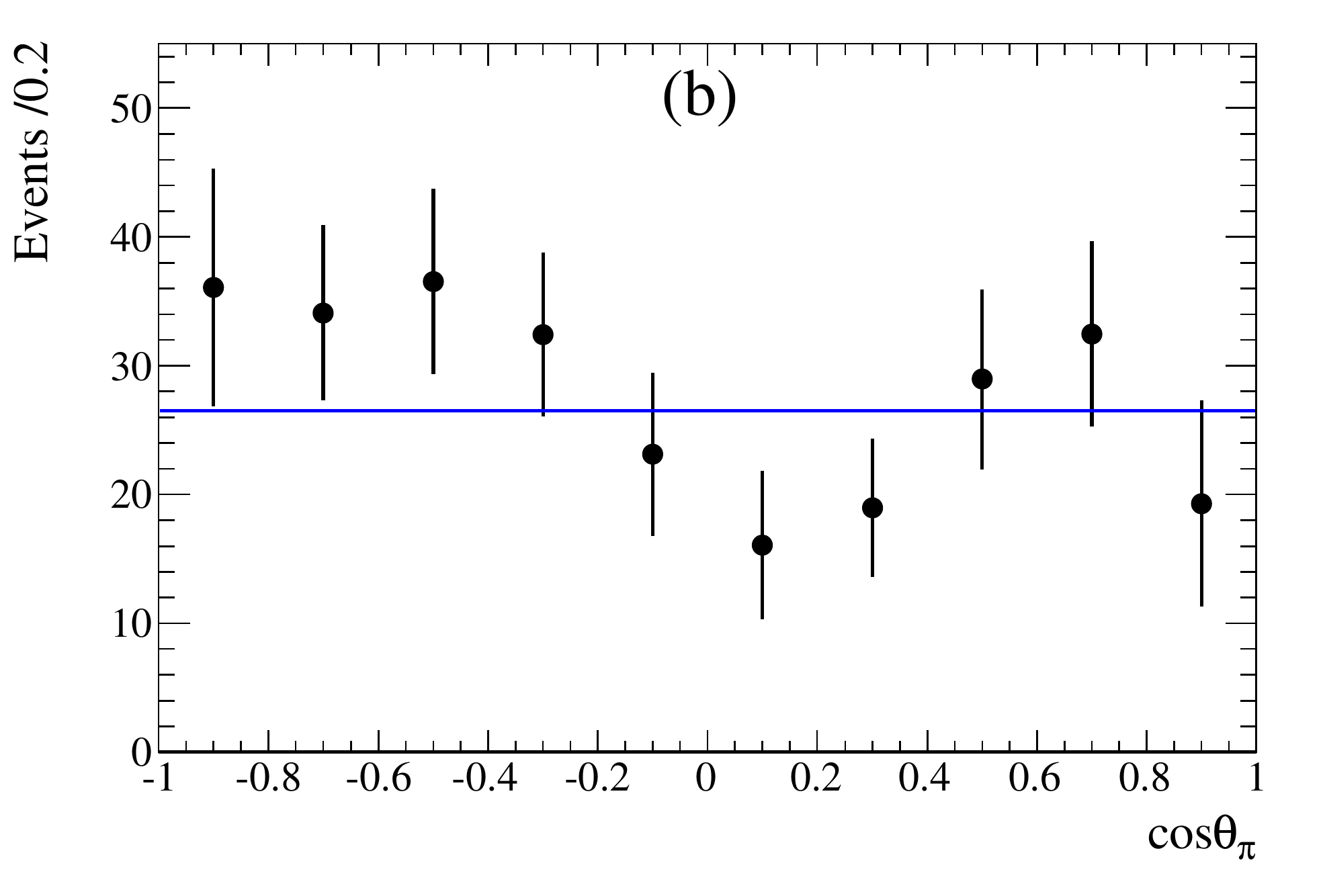}
	    \includegraphics[width=7.8cm,height=6cm]{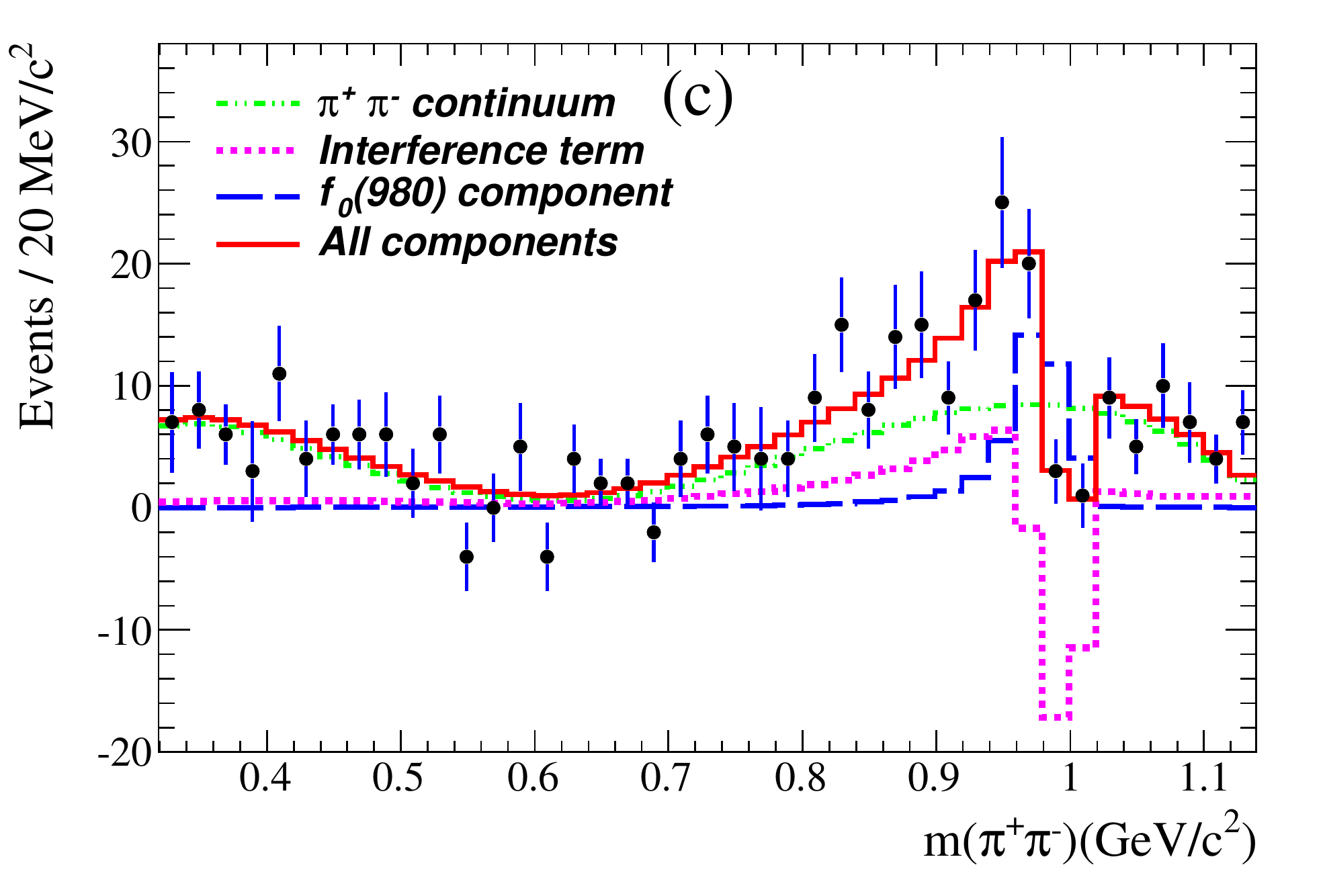}\\	
     \caption{(a) The background-subtracted \pipi mass distribution for the Y(4260) signal region; the dashed vertical line is at the nominal $f_{0}(980)$ mass value~\cite{ct:PDG}; (b) the corresponding cos$\uptheta_{\mathrm{\pi}}$ distribution; the fitted line is for an $S$-wave description; (c) the result of the fit using the model of Eq.~(\ref{f0int}).}
  \label{fig:pipimass}
\end{figure}
\indent We now consider the $\pipi$ system from Y(4260) decay to $\jpsi\pipi$. Since the Y(4260) has $I(J^{PC})=0(1^{--})$ and its width indicates strong decay, the \pipi system has $I(J^{PC})=0(0^{++})$ or $I(J^{PC})=0(2^{++})$. For the region 4.15 $\leq m(  \jpsi\pipi) \leq$ 4.45 \gevcc, the $\pipi$ mass distribution after subtraction of that from the \jpsi sideband regions is shown in  Fig.~\ref{fig:pipimass}(a). The region below 0.32 \gevcc is excluded since it is severely depopulated by the procedure used to remove $e^{+}e^{-}$ pair contamination. The distribution decreases from threshold to near zero at $\sim$ 0.6 \gevcc, rises steadily to a maximum at $\sim$ 0.95 \gevcc, decreases rapidly to near zero again at $\sim$ 1 \gevcc, and increases thereafter. The distribution is consistent with previous measurements~\cite{:2007sj,ct:babar-Y}.\\
\indent  We define $\uptheta_{\mathrm{\pi}}$ as the angle between the $\pi^{+}$ direction and that of the recoil  \jpsi, both in the dipion rest frame. The distribution in cos$\uptheta_{\mathrm{\pi}}$ is shown in Fig.~\ref{fig:pipimass}(b). The fitted line represents $S$-wave decay, and provides an adequate description of the data ($\chi^{2}/NDF=12.3 /9$, probability~=~19.7\%); there is no need for a $D$-wave contribution, $e.g.$, from $f_{2}(1270) \rightarrow \pipi$ decay.\\
\indent The mass distribution near 1 \gevcc suggests coherent addition of a nonresonant \pipi amplitude and a resonant amplitude describing the $f_{0}(980)$. If the peak near 950 \mevcc is attributed to a nonresonant amplitude with phase near $90^{0}$, the coherent addition of the resonant $f_{0}(980)$ amplitude, in the context of elastic unitarity, could result in the observed behavior, which is  similar to that of the $I=0$ \pipi elastic scattering cross section near 1 GeV (Fig.~2, p.VII.38, of Ref.~\cite{PDG1992}). However, we have no phase information with which to support this conjecture.\\
\indent The distribution in Fig.~\ref{fig:pipimass}(a) for $m_{\pi \pi} <0.9$ \gevcc is qualitatively similar to that observed for the decay $\Upsilon(3S)\to \Upsilon(1S)\pi^{+}\pi^{-}$~\cite{CLEOpipi}. There, the dipion mass distribution decreases from a maximum near threshold to a significantly non-zero minimum at $\sim$ 0.6-0.7 \gevcc, before rising steeply toward 0.8 \gevcc before being cut-off by the kinematic limit (0.895  \gevcc). The CLEO data are well-described in terms of a QCD multipole expansion~\cite{chan,yahn} up to $m_{\pi \pi}~\sim$0.7~\gevcc, but the sharp rise thereafter is not well-accommodated. This shortcoming is more readily apparent for the much larger~\babar~data sample for this same process~\cite{elisag}. There the distribution begins a rapid rise toward the $f_{0}(980)$ region, as in Fig.~\ref{fig:pipimass}(a), but turns over at $\sim$~0.85 because of the kinematic limit at 0.895  \gevcc. The CLEO multipole expansion fit involves two amplitudes whose relative phase ($\pm~155$ degrees) causes destructive interference, and hence the minimum in the mass distribution at $\sim$0.6-0.7 \gevcc. The amplitudes are of similar magnitude in this region, and so a relative phase of $\pm$~180 degrees could yield near--zero intensity, as observed in Fig.~\ref{fig:pipimass}(a). This phase value would result in an approximately real amplitude. However it would contain no explicit  $f_{0}(980)$ contribution, which seems necessary to a description of the data of Fig.~\ref{fig:pipimass}(a), and so we attempt to describe the entire distribution using the following simple model.\\
\indent The nonresonant intensity distribution requires three turning points, as in the CLEO multipole expansion description, and so we choose to represent it by a fourth-order polynomial, $T(m_{\pi \pi})$, where $m_{\pi \pi}$ is the invariant mass of the $\pi^{+} \pi^{-}$ system. From the phase requirement discussed above, it follows that the corresponding amplitude can be chosen to be real and represented by $ \sqrt{T(m_{\pi \pi}})$. To this amplitude we add the complex $S$-wave $\pi^{+} \pi^{-}$ amplitude obtained from the ~\babar~ analysis of $D_{s}^{+} \rightarrow \pi^{+}\pi^{-}\pi^{+}$ decay~\cite{antimo}, which shows clear resonant behavior at the $f_{0}(980)$. We perform a $\chi^{2}$--fit to the data of Fig.~\ref{fig:pipimass}(a) using \begin{equation}\label{f0int}      
f(m_{\pi \pi})=|\sqrt{T(m_{\pi \pi})} +e^{i\phi}F_{f_{0}(980)}(m_{\pi \pi})|^{2}\cdot p\cdot q~ ,
 \end{equation}
where $F_{f_{0}(980)}$ is proportional to the complex \pipi amplitude of Ref.~\cite{antimo}, and the phase $\phi$ is determined by the fit; $p$ is the $\pi^{+}$ momentum in the \pipi rest frame, and $q$ is the \jpsi momentum in the $\jpsi\pipi$ rest frame. We use the fitted Y(4260) mass value in calculating $q$, which implies a kinematic limit of 1.15 \gevcc for the fit function. The result is shown in Fig.~\ref{fig:pipimass}(c). The fit is good ($\chi^{2}/NDF=33.6 /35$, probability~=~53.6\%), and the interference contribution is important for the description of the region near 1 \gevcc ($\phi=28^{0} \pm 24^{0}$). The $f_{0}(980)$ amplitude squared gives $0.17  \pm  0.13 $~(stat) for the branching ratio ${\cal B} (J/\psi f_{0}(980), f_{0}(980)\to \pi^{+}\pi^{-})/ {\cal B}(J/\psi \pi^{+}\pi^{-})$. This is somewhat smaller than the prediction of Ref.~\cite{oset}, where it is proposed that the $f_{0}(980)$ contribution should be dominant. \\
\indent In summary, we have used ISR events to study the process  $e^+e^-\to\jpsi\pipi$ in the c.m.\ energy range 3.74--5.50 GeV. For the Y(4260) we obtain $m_Y=4244~\pm5$~(stat)~$\pm$4~(syst)$~\mevcc$, $\Gamma_Y=114~^{+16}_{-15}$~(stat)~$\pm7$~(syst)~$\mev$, and $\Gamma_{e^+e^-}\times{\cal B}(J/\psi\pi^{+}\pi^{-})  =9.2~\pm~0.8$~(stat)$~\pm0.7 $~(syst)~eV. 
These results represent an improvement in statistical precision of $\sim$ 30\% over the previous~\babar~results~\cite{ct:babar-Y}, and agree very well in magnitude and statistical precision with the results of the Belle fit which uses a single Breit-Wigner resonance to describe the data \cite{:2007sj}. We do not confirm the broad enhancement at 4.01\gevcc reported in Ref.~\cite{:2007sj}. The dipion system for the Y(4260) decay is in a predominantly $S$-wave state. The mass distribution exhibits an $f_{0}(980)$ signal, for which a simple model indicates a branching ratio with respect to $\jpsi\pipi$ of 0.17 $\pm$ 0.13 ~(stat).\\
\indent  We are grateful for the excellent luminosity and machine conditions
provided by our \pep2\ colleagues, 
and for the substantial dedicated effort from
the computing organizations that support \babar.
The collaborating institutions wish to thank 
SLAC for its support and kind hospitality. 
This work is supported by
DOE
and NSF (USA),
NSERC (Canada),
CEA and
CNRS-IN2P3
(France),
BMBF and DFG
(Germany),
INFN (Italy),
FOM (The Netherlands),
NFR (Norway),
MES (Russia),
MICIIN (Spain),
STFC (United Kingdom). 
Individuals have received support from the
Marie Curie EIF (European Union)
and the A.~P.~Sloan Foundation (USA).


\bibliography{paper}

\begin{thebibliography}{99}

\bibitem{ct:X3872} 
S.~K.~Choi {\it et al.},
\jprl{91}, 262001 (2003).

\bibitem{ct:Z3930}
S. Uehara {\it et al.}, 
\jprl{96}, 082003 (2006).

\bibitem{ct:Y3940}
S.~K.~Choi {\it et al.}, 
\jprl{94}, 182002 (2005).

\bibitem{ct:X3940}
K. Abe {\it et al.}, 
\jprl{98}, 082001 (2007).

\bibitem{ct:charmonia}
T. Appelquist, R.~M. Barnett and K.~D. Lane, Ann. Rev. Nucl. Part. Sci. {\bf 28}, 387 (1978).

\bibitem{ct:babar-Y} 
B.~Aubert {\it et al.},
\jprl{95}, 142001 (2005).

\bibitem{Coan:2006rv}
  T.~E.~Coan {\it et al.}  
  Phys.\ Rev.\ Lett.\  {\bf 96}, 162003 (2006).

\bibitem{ct:ddstar} 
B.~Aubert {\it et al.},
\jprd{76}, 111105 (2007); 
B.~Aubert {\it et al.}, \jprd{79}, 092001 (2009); 
G. Pakhlova {\it et al.},  \jprl{98}, 092001 (2007); 
G. Pakhlova {\it et al.},  \jprd{77}, 011103 (2008).

\bibitem{ct:dsdstar} 
P. del Amo Sanchez {\it et al.}, \jprd{82}, 052004 (2010); 
J. Libby {\it et al.}, Nucl. Phys. B, Proc. Suppl. {\bf 181-182}, 127 (2008).


\bibitem{ct:tetra}
I. Bigi {\it et al.},
\jprd{72}, 114016 (2005).

\bibitem{ct:tetra2}
L. Maiani {\it et al.},
\jprd{72}, 031502 (2005).


\bibitem{ct:baryonium}
C.~-F. Qiao,
J. Phys. G {\bf 35}, 075008 (2008).

\bibitem{ct:hybrid}
S.~-L.~Zhu, Phys. Lett. B {\bf 625}, 212 (2005).

\bibitem{:2007sj}
C.~Z.~Yuan {\it et al.},
\jprl{99}, 182004 (2007).

\bibitem{Aubert:2006ge}
B.~Aubert {\it et al.},
\jprl{98}, 212001 (2007).

\bibitem{:2007ea}
X.~L.~Wang {\it et al.},
\jprl{99}, 142002 (2007).

\bibitem{ct:babar-detector}
B.~Aubert {\it et al.}, 
\nima{479}, 1 (2002); W. Menges, IEEE Nucl. Sci. Symp. Conf. Rec. 5, 1470 (2006).

\bibitem{ct:PDG}
K. Nakamura  {\em et al.}, J. Phys. G {\bf 37}, 075021 (2010), and partial update for the 2012 edition (URL:http://pdg.lbl.gov).

\bibitem{MC}
  G. Bonneau and F. Martin, Nucl. Phys. {\bf B27}, 381 (1971);
 H.Czyz {\it et al.}, Eur. Phys. J. {\bf C18}, 497 (2001); D. J. Lange, Nucl. Instrum. Methods Phys. Res., Sect. A {\bf 462}, 152 (2001).

\bibitem{GEANT}
  S. Agostinelli  {\it et al.}, Nucl. Instrum. Methods Phys. Res., Sect A{\bf 506}, 250 (2003).


\bibitem{taumass}
B.~Aubert {\it et al.}, \jprd{80}, 092005 (2009). 

\bibitem{lambdamass}
B.~Aubert {\it et al.}, \jprd{72}, 052006 (2005). 


\bibitem{Kuraev:1985hb}
  E.~A.~Kuraev and V.~S.~Fadin,
  Sov.\ J.\ Nucl.\ Phys.\  {\bf 41}, 466 (1985).

\bibitem{Adam:2005mr}
  N.~E.~Adam {\it et al.} ,
  Phys.\ Rev.\ Lett.\  {\bf 96}, 082004 (2006).
     
\bibitem{BES}
  J.~Z.~Bai {\it et al.},
  Phys.\ Lett.\  B {\bf 605}, 63 (2005).


\bibitem{PDG1992}
K. Hikasa {\it et al.}  \jprd{45}, Part II (1992).

\bibitem{CLEOpipi}
D. Cronin-Hennessy  {\it et al.}   \jprd{76}, 072001 (2007).

\bibitem{chan}
L.~S. Brown and R. N. Cahn,  Phys.\ Rev.\ Lett.\  {\bf 35}, 1 (1975).

\bibitem{yahn}
T.~-M. Yahn, \jprd{22}, 1652 (1980).

\bibitem{elisag}
E. Guido, Ph.D. Thesis, Universit\'a degli Studi di Genova (2011);
~\babar~ Publications, THESIS-12/003 (2012).


\bibitem{antimo}
B.~Aubert {\it et al.}, \jprd{79}, 032003 (2009).

\bibitem{oset}
  A.~Martinez Torres  {\it et al.},
  Phys.\ Rev.\  D {\bf 80}, 094012 (2009).


\end{thebibliography}

\end{document}